Viewpoint

# Is a gene-centric human proteome project the best way for proteomics to serve biology ?

Thierry Rabilloud [1], Denis Hochstrasser [2] Richard J Simpson [3]

1 Biochemistry and Biophysics of Integrated Systems, UMR CNRS-CEA-UJF 5092, CEA Grenoble, iRTSV/BSBBSI, 17 rue des martyrs, F-38054 GRENOBLE CEDEX 9

2 Department of Genetic & Laboratory Medicine, Laboratory Medicine Division, Geneva University, Geneva University Hospital & Swiss Centre of Applied Human Toxicology
Rue Gabrielle-Perret-Gentil, 4
CH-1211 Geneva 14, Switzerland

3 Ludwig Institute for Cancer Research Inc
PO Box 2008, Royal Melbourne Hospital,
Royal Parade, Parkville, Victoria, Australia 3050

Correspondence to

Thierry Rabilloud, iRTSV/BBSI
CEA-Grenoble, 17 rue des martyrs,
F-38054 GRENOBLE CEDEX 9
Tel (33)-4-38-78-32-12
Fax (33)-4-38-78-44-99
e-mail: Thierry.Rabilloud@ cea.fr

Abstract

With the recent developments in proteomic technologies, a complete human proteome project appears feasible for the first time. However, there is still debate as to how it should be designed and what it should encompass. In 'proteomics speak', the debate revolves around the central question as to whether a gene-centric or a protein-centric proteomics approach is the most appropriate way forward. In this paper, we try to shed light on what these definitions mean, how large-scale proteomics such as a human proteome project can insert into the larger omics chorus, and what we can reasonably expect from a human proteome project in the way it has been proposed so far.

In 2010, the tenth anniversary of the completion of the Human genome project is being celebrated [1]. Even if this project has not met all expectations in terms of impact on human health, it is fair to say that genomics-derived techniques, especially expression arrays [2] and deep sequencing, are now very widely used and are having an impact in many fields, going from basic biology to clinical and toxicological studies. Moreover, sequencing speed has dramatically increased over this decade, and close to 4000 organisms have now had their genome sequenced[3].

By contrast, in the proteomics field, the first comprehensive proteome, i.e. at least a complete list of expressed proteins in a defined condition, is yet to materialise, although recent technological developments have made this goal more realistic [4]. While the idea of a repertoire of the human proteins has preceded by far even the dream of a human genome project and the coining of the word proteomics [5], there has been much recent debate on the scope of a Human Proteome Project (HPP) [6-9], and the launch of this initiative has been recently announced [10].

In light of this gap in maturity between genomics and transcriptomics on the one side, and proteomics on the other, we feel that the proteomics community needs to step back and reflect upon (i) where and how a HPP, compared to other omics, would add value to the current challenge for biology, i.e. determining gene product functionality and how proteins interact to make a viable organism, and (ii) whether a HPP in its current shape, i.e. gene-centric, is the best way for proteomics to deliver on this important challenge. This utility-over-resource debate is still pending in genomics [2] [11], and it is quite clear that proteomics cannot escape it, although it is quite tempting to think that proteomics will also play an indispensable role in addressing the major challenge of selecting the few true-positive associations with disease in Genome-Wide Association Studies [12]

Prior to entering the debate, it is appropriate to better define how proteomics is made, and to comment on the strengths and weaknesses of each of the two major types of proteomics.

The first type of proteomics, defined as protein-centric [7] or discovery-oriented, consists in identifying as many proteins as possible in complex biological matrices (tissues, body fluids, cellular organelles etc). To this purpose, highly efficient protein separation tools such as 2D gels and/or protein/peptide electrophoretic or chromatographic separations coupled to highly-intensive MS methods are used [13, 14]. In other words, we try to observe in an unbiased manner what nature allows us to observe, the important advantage being that we can observe unpredicted protein forms, - e.g., post-translationally- modified peptides, alternate splice variants, proteolytically processed low-Mr peptides etc. In the current state of the art, several thousands of proteins can be identified in a complex sample [15, 16], although the details per protein decrease at such numbers.
In this scheme, the price to pay is that if a protein is not observed, we never know if this is due to the fact that it is really not present or that we are not able to detect it due either to the dynamic range constraints of the MS hardware / protein separation technologies or to physico-chemical features of proteins or peptides that prevent them from being efficiently analyzed.

The second type of proteomics, defined as gene-centric or scoring proteomics, starts from a completely different paradigm. Bluntly speaking, the question shifts from a broad "what proteins are in my sample? " to "is protein X present in my sample, and if yes, at what level ?" Answering this question means to be able to detect and quantify unequivocally each protein in a complex sample. To this purpose, the information contained in sequences database about each of the ca. 20,000 human protein-coding genes is used to derive proteotypic peptides, i.e. unique peptides specific for an individual protein. These proteotypic peptides can be used in two different analytical setups that impact on how proteotypic peptides are defined and selected.

In the first, antibody-based setup [17], protein fragments containing the proteotypic peptides, or recombinant proteins, are used to raise antibodies, the specificity of which is then tested. In this case, immunogenicity is an important parameter in selecting the protein pieces used as the starting material.

In the second, MS-based setup, proteotypic peptides are analyzed with a specialized MS mode, called multiple reaction monitoring (MRM), also referred to as single reaction monitoring (or SRM) experiments [18, 19]. In this case, tryptic peptides will be selected, and apart from the specificity, performance of the peptide in MS will be a critical parameter [20, 21]. In this regard, databases of proteotypic peptides are very helpful [22].

The enormous advantage of the gene-centric format is that many key parameters can be characterized upfront, such as lower detection limits, lower quantification limits, precision, accuracy etc., even in the complex sample of interest, just by spiking known quantities of your observable (immunogen in the antibody format and isotopically-labelled proteotypic peptide in the MRM format). Thus, when you get a negative response, you can translate it into a threshold of detection, exactly as in the transcriptome arrays. Furthermore, as you simplify the proteome into gene-products, the precise (almost) number you know, the task is circumscribed to a limited (although high) number of measurements, rather than being without predictable limits, as in the case of protein-centric proteomics.

Moreover, this scoring proteomics, both with antibodies and SRM, is scalable. It can be used to quantitatively monitor hundreds of known proteins (say, all known components of multiple signaling pathways involved in cancer) simultaneously, but it can be envisioned to expand it at a genome-wide scale, here again both for the antibody format [17] and for the SRM format [23], and this explain why the gene-centric format has been chosen by HUPO for conducting the HPP [10].

As the HPP will be a "billion-dollar baby", the key is to figure out what can, or cannot, be achieved. In this sense, it may be interesting to thoroughly mine extant data that the broader proteomics community has amassed, collectively, over this past decade. If these data were (i) collected in a few standardized formats in repositories and (ii) evaluated by an international consortium, using robust metrics of data validation, to separate the grain from the chaff, we would know – for relatively little cost – just how many gene products (proteins – including post-translational modifications and alternate splice variants) we have vaulted away already. Surely, it is incumbent on the proteomics community to establish the *status quo* before we embark on a highly-expensive HPP.

However, it must be kept in mind that there is still a major gap in comprehensiveness between proteomics, on the one hand, and genomics and transcriptomics, on the other hand. Just as an example of how this gap impacts a scientific field, large-scale toxicological analysis is now almost exclusively done through cDNA arrays (i.e. by transcriptomics) despite the initial successes demonstrated by proteomics in this field and its higher relevance (e.g. in [24] ). This gap in comprehensiveness is linked to built-in features, such as the much higher dynamic range in protein concentrations [25, 26] or the chemical heterogeneity and instability of proteins.

Thus, it is safe to forecast that proteomics will be used mostly in areas where only proteomics can deliver, and not the other, cheaper and more comprehensive *omics*.

The first, obvious area is to obtain evidence of the presence and of the quantity of a gene product directly at the protein level. Indeed, a fair number of predicted gene products have not been observed yet at the protein level, and this type of use is the main current use of proteomics, especially for discovery-oriented proteomics.
This area, however, will be the one where the competition with genomics-derived techniques will be most felt, especially with the issue of comprehensiveness. Although genomics-derived methods provide only indirect evidence of protein expression, they have long demonstrated their comprehensiveness. Oppositely, recent efforts for genome-wide proteomics in yeast [4] strongly suggest that comprehensiveness may be difficult to reach in proteomics. In this work, the success rate for SRM in yeast is claimed to be 90%. This means in turn that even in yeast, 10% of the proteins will be difficult to bring to analysis, and for the moment, we have no idea of how much it will cost to reach comprehensiveness in yeast. Transposed to the human proteome, that grounds on 20,000 genes instead of 6000 in yeast and comprises more complex proteins families (for which proteotypic peptide are more difficult to find to differentiate each member of the family), the success rate can be anywhere between 70 and 90%, and the cost for the asymptotical reach for comprehensiveness is unpredictable. This key issue of the cost of comprehensiveness, compared to the power of existing genomics-derived technologies, makes this area very competitive for the HPP.
Consequently, a HPP will be much more valuable in areas where genomics cannot deliver easily,

The first of these areas is clearly obtaining information on protein localization. A typical example occurs when subcellular localizations are investigated. Proteins are made essentially in the cytosol, but many are then relocated to various cellular sites (membrane, nucleus, organelles) where their play their roles. As exemplified by the prenylation of the small G-proteins, such a relocalization event can be crucial for the activity of the protein [27].

For this question, antibodies have long proven that they are a format of choice to investigate this problem, especially now with tools such as confocal microscopy. The other, mass spectrometry-based formats are more difficult to use (although they are quite extensively used nowadays in discovery-oriented proteomics) because they must rely on biochemical fractionation of the sample. This process is usually not very efficient in terms of purity, and the contamination issue must be addressed by labor-intensive methods such as a posteriori verification by antibodies [28] or protein correlation profiling [29].
The second interesting situation is represented by tissue leakage, and observation of tissue-specific proteins in body fluids. This situation is amply used in current clinical tests (e.g. the very classical transaminase measurements), and is a major predictable application area for proteomics. In this case, years of discovery-oriented proteomics have failed to detect tissue leakage proteins in body fluids for quite clear reasons [30], and this precise aspect will be easier to tackle with a gene-centric proteomics format (either antibody or MS-centric) [31]. However, protein degradation or modification could be the hallmark of some disease processes and this would require a protein-centric proteomic approach to be discovered.

Obtaining evidence of protein-protein interactions is also a key area in a HPP, and has always been advocated in all HPP proposals. In this topic, the composition of ex vivo-purified protein complexes is determined, and complex protein interaction graphs can then be drawn [32]. Generally speaking, the complexes are of moderate complexity (a ribosome is less than 100 different proteins) and dynamic range (stoichiometry is limited) so that protein-centric formats are usually used for this type of proteomics [33]. However, when examining the utility of proteomics for the biology community at large, the key question is the reality of the interactions that can be evidenced. In all cases there is one intractable problem, i.e. the possible scrambling of protein complexes during the preparation of a workable extract. But there are other, format-depending problems.

In pull-downs experiments using antibodies, the artifacts may arise mainly from the fact that this format operates in vast excess of reactants (antibodies and beads) over analyte (the complex of interest). The problem is that both antibodies and beads are able to interact with some proteins independently of the antigen-binding sites [34], so that there is often a bucket of artefactual proteins for a spoonful of complex of interest, leading to serious specificity and analytical problems.
In pull-downs by tandem affinity methods, as first exemplified by the classical TAPTAG approach [35], specificity and quantitative ratios are usually better. However, there is a key point often forgotten when such approaches are used in mammalian cells. In their first applications in yeast [35], the baits used for fishing of complexes were stably introduced in yeast by homologous recombination- i.e., in place of the endogenous gene and under the control of the endogenous promoter. In mammalian systems, the bait is usually introduced transiently by a retroviral vector [36], i.e., under the control of the strong LTR promoter. This means in turn that the bait is expressed at levels that may be quite different from those of the endogenous protein, which moreover is still present. Thus, the core assumption that the complex assembling on the bait is the same as the one assembling *in vivo* may well be wrong. For example, the almost systematic presence of chaperones in mammalian complexes may well be a reaction to protein over-expression, and not representative of the real in vivo situation.

Deciphering the role of post-translational modifications is however the area where proteomics can have the most impact in biology. In fact, recent advances in genomics bring to light a very important question: how can we be made of more than 300 different cell types, assembling in so complex patterns to build complex but defined organs, with "only" ca. 20,000 genes? In other words, the complexity that is not present at the gene level must present somewhere else, and this somewhere is at the protein level, in two ways. The first way to create complexity is the law of mass action, by which changes in protein concentrations will alter metabolic fluxes, but also protein-protein associations and thus the building of complexes and superstructures. The second way is to introduce chemical diversity beyond the genome by modifying the proteins at the post-translational level. Of course, the induced changes in protein chemical properties can bring changes into every function of the protein, from catalytic activity (e.g. kinases) or cellular localization (e.g. prenylation of small G proteins, phosphorylation of some transcription factors) [37] to protein degradation (phosphorylation of IKKb) or protein incorporation into complexes. Thus, there is ample evidence that understanding how post-translational modifications modulate protein functionalities will provide key understanding of the molecular processes of life.

In this area, it is quite clear that gene-centric proteomics cannot be expected to perform well. First of all, gene-centric proteomics is a targeted technique, which therefore needs *prior* knowledge of the modification to look at [38]. This knowledge is obviously far from being available now, as shown by new modifications arising in the landscape (e.g., propionylation in addition to acetylation [39] ). Moreover, in the MRM format for gene-centric proteomics, and despite the ever increasing knowledge on how to predict "useful" peptides for this type of assay [21, 40, 41], not every proteotypic peptide can be translated into a useful MRM assay, as shown in the yeast example [4] because of the structural constraints brought by the MS process itself [41] .

Consequently, this area of post-translational modifications will remain an area for protein-centric proteomics for quite some time. Moreover, there is an area that needs to be seriously explored now, which is the cooperativity (or anticooperativity) between modifications on proteins. There is the well-known example of phosphorylation vs monoglycosylation (e.g. in [42]), but this is a simple case of competition of modifications at the same site. More generally speaking, how modification at site X promotes or hampers modification at site Y on other proteins is still largely unknown, although we begin to have a flavor of these phenomena on some very well-studied proteins, such as the antioncogenic protein p53 [43, 44]. This relative lack of knowledge is largely due to the fact that what we call protein-centric proteomics is more and more peptide-based (shotgun) rather than protein-based (2D gels), while an integrated knowledge of protein modifications requires working with intact proteins rather than peptides alone [45, 46].

Moreover, in clinical proteomics, it should be kept in mind that the origin of diseases could be simplified in two large categories: the genes and the environment. The environment can be further subdivided into microbes and toxics in a large sense. Toxics modify proteins at many levels, some unrelated to gene mutation or expression (e.g. in [47]). Therefore, only a protein-centric approach can discover some of the environmentally produced modifications found at the peptide level.

When summarizing the whole landscape, rather simple guidelines emerge. Shifting from protein-centric to gene-centric proteomics equates to bartering details for comprehensiveness. As all the other omics are comprehensive, we probably must consider this aspect. However, when going down in complexity, and dealing with more and more purified samples, which are both less complex in protein numbers and in dynamic ranges, there is clearly a threshold where protein-centric proteomics can do the job in terms of comprehensiveness, with the added bonus of many modifications brought to light. Of course this cursor of the best choice between gene-centric and protein-centric approaches may change its position in the future with the improvement of both gene-centric and protein-centric proteomic techniques.

However, it must be kept in mind that protein-centric proteomics of subcellular fractions is a field where proteomics has already had an important impact in the corresponding biology area, as demonstrated by the examples of the nuclear pore [28] , centrosome [48] or exosome [49]. This clearly shows the great value of protein-centric approaches in key areas of the basic cell biology and biochemistry, as well as in many areas of clinical biology, where it becomes clear that molecular diagnosis in general would be improved by taking into account post-translational modifications of marker proteins (e.g. in [50]).

In conclusion, we would like to draw a parallel between proteomics and archeology, another way of bringing hidden treasures to light. Quite often in archeology, you can see by an aerial photograph the hidden roman villa that is buried in a field. But then you must go there, dig it out, and in the end use a small paintbrush to exhume the finest details – i.e., details that provide the uniqueness and value to your effort. Thus, in proteomics too, we do not want to have to choose between the plane and the paintbrush, between the gene-centric and the protein-centric approaches, between scoring proteomics and discovery proteomics, because we know that we need both to have a real impact in biology.